\documentclass{ametsocV6.1}
\usepackage{float}
\usepackage{soul}
\usepackage{natbib}

\title{The Impact of Cumulus Clouds and CCNs Regeneration on Aerosol Vertical Distribution and Size }

\authors{Yael Arieli\aff{a}, Alexander Khain\aff{b}\correspondingauthor{Alexander Khain, alexander.khain@mail.huji.ac.il}, Ehud Gavze\aff{b}, Orit Altaratz\aff{a}, Eshkol Eytan\aff{a},  Ilan Koren\aff{a}\correspondingauthor{Ilan Koren, Ilan.Koren@weizmann.ac.il}}

\affiliation{\aff{a}{Department of Earth and Planetary Science, Weizmann Institute of Science, Rehovot, Israel}
\aff{b}{Institute of Earth Science, Hebrew University, Jerusalem, Israel}}

\abstract{This study employs a high-resolution (10m) System for Atmospheric Modeling (SAM) coupled with the Spectral Bin Microphysical (SBM) scheme to thoroughly investigate the processes governing the evolution of aerosol properties within and outside a shallow cumulus cloud. The model encompasses the complete life cycle of cloud droplets, starting from their formation through their evolution until their complete evaporation or sedimentation to the ground. Additionally, the model tracks the aerosols' evolution both within droplets and in the air. Aerosols are transported within the droplets, grow by droplet coalescence, and are released into the atmosphere after droplet evaporation (regeneration process). The aerosol concentration increases by droplet evaporation and decreases along with falling drops. So, the effects of clouds on the surrounding aerosols depend on the microphysical and dynamic processes, which in turn depend on the amount of background aerosols; here, we compare clean and polluted conditions. 
It is shown that clouds significantly impact the vertical profile of aerosol concentration in the lower troposphere, as well as their size distribution, and can serve as a source of large (giant) cloud condensation nuclei.  
Furthermore, it is shown that both precipitating and non-precipitating boundary layer clouds contribute to a substantial increase in aerosol concentration within the inversion layer due to intense evaporation.}

\begin{document}

%% Necessary!
\maketitle

\section{Introduction}

Clouds are a major component in the climate system, generally covering half to two-thirds of the globe. Clouds reflect incoming radiation and absorb outgoing longwave radiation, thus exerting both cooling and warming effects on our planet (\cite{hartmann1992effect, ramanathan1989cloud,SensitivityofRadiativeForcingtoVariableCloudandMoisture, CloudFeedbackProcessesinaGeneralCirculationModel}). Moreover, they are one of the controlling factors in regulating the troposphere's water vapor content, an important greenhouse gas (\cite{cess1975global,ramanathan2006radiative,sun1993distribution}).

Aerosol particles also have a major climatic radiative effect. They interact with solar radiation through absorption and scattering and, to a lesser extent, with terrestrial radiation through absorption, scattering, and emission. In addition, aerosol particles play a fundamental role in cloud formation and evolution as they act as cloud condensation nuclei (CCN). The activation of CCNs means the formation of cloud droplets. Therefore, CCNs' concentration, size, and properties are key parameters in cloud micro-scale and macro-scale processes and properties, such as cloud radiative properties (\cite{albrecht1989aerosols,twomey1977influence}), precipitation development (\cite{squires1958microstructure}), and the turbulent mixing of clouds with their environment (\cite{bretherton2007cloud}). 
Many studies explored the topic of aerosol-cloud interactions, focusing on how changes in the aerosol amount and properties affect clouds' microphysics and dynamics (\cite{altaratz2014cloud,khain2009notes,khain_book}). However, the influence of clouds on the concentration and size distribution of aerosol particles, and particularly the role of cumulus clouds in the transport and alteration of the aerosol distribution within the surrounding atmosphere, is often overlooked in cloud simulations. Understanding how clouds modify the aerosol size distribution is crucial due to its implications for the next clouds that develop in the field. The region around clouds, which is the transition between clouds and clear skies, known as the Twilight zone, contributes significantly to the atmospheric radiative properties (\cite{eytan2020longwave,koren2009aerosol,jahani2022longwave}). This zone is occupied by liquid droplets and haze (humidified aerosols). Observations from satellites and ground-based sources have indicated its existence up to approximately 30 kilometers away from clouds, exhibiting an e-fold behavior over a 10-kilometer distance (\cite{koren2007twilight}).

The efficiency of an aerosol particle in acting as a CCN is determined by its size and hygroscopicity (chemical composition). Particles with larger sizes and higher hygroscopicity are more efficient in forming cloud droplets, thereby requiring lower supersaturation conditions. Understanding the temporal and spatial variability of aerosol particle concentration is crucial due to the multifaceted roles these particles play, including their impact on cloud formation and processes. 

Soluble aerosols appear as haze particles in humid air ($RH>\sim70\%$).  The size of a haze particle can be several times larger than that of a dry aerosol. Complete droplet evaporation would result in haze particle formation, according to the environmental relative humidity. Its radiative properties depend on the particle size.  In large-scale models, the optical properties of aerosols depend on air humidity, meaning they are treated as haze (\cite{morcrette2009aerosol,tang1997thermodynamic,tang1994aerosol}).

The regeneration is a mechanism of aerosol formation due to complete drop evaporation. Previous studies have considered this mechanism, but in a simplified manner, often utilizing 2D models or moment-based calculations of aerosol size distributions (\cite{flossmann1985theoretical,feingold1996numerical,xue2010effects}). 
\cite{fan2009ice} used a 3D model and assumed that drop evaporation leads to a release of aerosols with a size distribution proportional to the initial background aerosols distribution. \cite{shpund2019simulating} simulated a mesoscale convective system using WRF-SBM. In that study, the parameterization of aerosol return was similar to that in \cite{fan2009ice},  but the drop evaporation led to the release of aerosol distribution proportional to the distribution of activated aerosols at each grid point.  In \cite{magaritz2010effects} study, the effects of a stratocumulus cloud on aerosols were simulated using a 2D Lagrangian-Eulerian bin microphysical model with aerosol within droplets. It was shown that collisions increase the aerosol size, but the drizzle formation largely eliminates the largest aerosols. \cite{lebo2011continuous} examined marine stratocumulus deck in a small domain. They developed two-dimensional aerosol-cloud microphysical models that predict the simultaneous development of the discretized aerosols and drop size distributions. This was achieved by integrating a bin aerosol model with a bin microphysical model and computing the transfer of aerosol solute mass between drops, relying on water mass transfer principles. They showed that this treatment predicts increased LWP with increased aerosol loading.  
\cite{hoffmann2023note} analyzed 3D simulated stratocumulus using a Lagrangian cloud model coupled with LES (\cite{hoffmann2015entrainment}) and showed that the aerosol size distribution is shifted toward larger sizes due to collision coalescence processing.
We are unaware of studies investigating small cumulus' effects on the environmental aerosol field.

In this study, we conducted high-resolution simulations of single cumulus clouds forming at different aerosol concentration conditions. By tracking the aerosol evolution within droplets and in the air, we calculate the changes in the aerosol size distribution in the domain, within the clouds, and around them.

\section{Methods}

\subsection{Model Description}
In this study, we employed the \textbf{S}ystem for \textbf{A}tmospheric \textbf{M}odeling (SAM) (\cite{khairoutdinov_2003}) coupled with the Spectral Bin Microphysical (SBM) scheme \cite{khain2004simulation,khain2008factors}.
The microphysical scheme solves equations for two size (number) distribution functions of water droplets and dry aerosols (serving as cloud condensation nuclei, CCN). They are calculated on two different logarithmic doubling mass grids, each containing 33 bins. The minimum radius size of the dry aerosol particles is $0.0012$-$\mu$m, and the maximal one is 2-$\mu$m.

Based on the Köhler theory and using the supersaturation values for water, the critical dry aerosol radius is calculated, and dry aerosols larger than this size are nucleated into droplets. The corresponding bins in the dry aerosol size distribution are emptied. Dry aerosols smaller than 0.33$\mu$m are nucleated into the smallest droplet size bin (2 $\mu$m), whereas the size of droplets forming on larger aerosols is determined by multiplying the dry aerosol size by a factor of 6 (following \citet{kogan1991simulation}). This factor is estimated based on detailed calculations of ascending parcels while considering that large CCN cannot reach haze equilibrium size. The smaller dry aerosols are advected with the air and can be activated at later times in case of increased supersaturation.

The diffusional growth and evaporation are calculated using a semi-analytical approach, solving a coupled system of differential equations to determine droplet growth and the corresponding decrease in supersaturation simultaneously.   
To decrease the droplet size distribution (DSD) artificial broadening, a method analogous to a movable mass grid (\cite{kogan1991simulation}), as well as modified remapping techniques, are employed. These approaches for decreasing the numerical diffusion allow the model to be sensitive to aerosol concentration and predict well the DSD width and the height of the first radar echo (\cite{benmoshe2012turbulent,khain2019parameterization}). 
 
Collision–coalescence is solved by the stochastic collision equation with minimal diffusivity based on exponential flux method following \cite{bott1998flux}. The collision kernels are calculated using an exact superposition method (considering the flow fields around colliding droplets) described by \cite{pinsky2001collision}. 
 
Sedimentation is calculated using fall velocities determined by \cite{beard1976terminal}. 

The new component we have implemented into the model is the tracking of aerosol evolution within the cloud droplets. This enables a description of the return of the dry aerosols to the atmosphere after droplets' complete evaporation. In the droplet nucleation process, the activated dry aerosols are transferred from the size distribution of the dry aerosol to the DSD grid.  In case dry aerosols are soluble, it actually means that the droplets represent a weak solution. However, the aerosol mass within the droplet is considered for the tracking. The aerosols in the droplets are described by a new (third) size distribution grid (containing 33 bins), reflecting the distribution of droplet salinity or dry aerosol mass fraction. We calculate the total aerosol mass in each drop size bin and the average aerosol mass per drop.
The changes in aerosol mass within the droplets are tracked and calculated according to the microphysical processes. During droplet diffusional growth (or partial evaporation), the droplet water mass increases (decreases), while the aerosol mass doesn't change. Drop-drop collisions and coalescence form drops whose mass equals the sum of the two drop masses. The aerosol mass in the resulting drop is the sum of aerosol masses in the colliding drop. Aerosols within drops are advected, mixed, and sediment with the corresponding drops.
% ===============================

Once a droplet completely evaporates, the aerosol shifts into the size distribution of dry aerosols. It moves into the bin that corresponds to its mass. It is assumed that the evaporation of one droplet results in the release of one aerosol particle. This phenomenon is referred to hereafter as 'regeneration'.  Subsequently, if a regenerated aerosol particle re-enters the cloud, it can re-activate and become a droplet (depending on the supersaturation conditions). 

It should be noted that the humidity conditions around the cloud support haze formation. At sub-saturation conditions, all haze particles can be assumed to be in equilibrium with the surroundings.  The radius of the haze particles ($r_{eq}$) is calculated using the Köhler theory:

\begin{equation}
  S = \frac{A}{r_{eq}}-\frac{Br_N^3}{r_{eq}^3-r_N^3} \\
\end{equation}
where S is the environmental saturation (as a fraction), $r_N$ is the aerosol radius, $r_{eq}$ is the haze radius at equilibrium, and \textit{A} and \textit{B} are coefficients weakly depending on temperature (eq. 5.1.11 from \cite{khain_book}). \\
\subsection{Simulation Setup}
 
Four single cumulus clouds (Cu) simulations were conducted using the BOMEX thermodynamic conditions. The setup was taken from \cite{ALargeEddySimulationIntercomparisonStudyofShallowCumulusConvection}, including the large-scale forcing, vertical profiles of water vapor mixing ratio and potential temperature, with an inversion layer located at $1500–2000 m$. The background wind was set to zero, and the surface fluxes were kept constant.
Following \cite{jaenicke1988aerosol} and \cite{altaratz2008aerosols}, the size distribution of the aerosols was prescribed as the sum of three log-normal distributions describing fine, accumulation, and coarse aerosols modes, where in these simulations, the fine mode was set to zero.
The clouds were initiated by a thermal perturbation of $0.1 K$ in two different aerosol concentration conditions. The first cloud developed in a clean environment (clean cloud, dry aerosol concentration of $50 cm^{-3}$), and the second cloud developed in polluted conditions (polluted cloud, dry aerosol concentration of $500 cm^{-3}$). The initial dry aerosol concentration was set as a constant number below the cloud base ($600 m$) and zero above this altitude since this study aims to investigate the aerosol field around the cloud as caused by droplet evaporation. All dry aerosol particles were assumed to be Ammonium sulfate and to serve as CCN. Two simulations were conducted for each type of cloud (clean or polluted, a total of 4 simulations). The first simulation included the regeneration process, while the second did not.  \\

The domain size was $5.12$ $km$ X $5.12$ $km$ X $4$ $km$ (which is substantially larger than the cloud size), with lateral cyclic boundary conditions.  The horizontal resolution was $dx=dy=10 m$ and the vertical resolution $dz=10 m$ up to $3 km$, and $50 m$ for the upper-most kilometer. The simulation's time step was $dt=0.5 sec$.

\section{Results}
\subsection{The Clouds' Effect on the Aerosol Vertical Profiles}
\begin{figure}[H]
  % \noindent\includegraphics[width=19pc,angle=0]{Nccn_crossction_32to52_50CCN_3.jpg}\\
  \noindent\includegraphics[width=\linewidth]{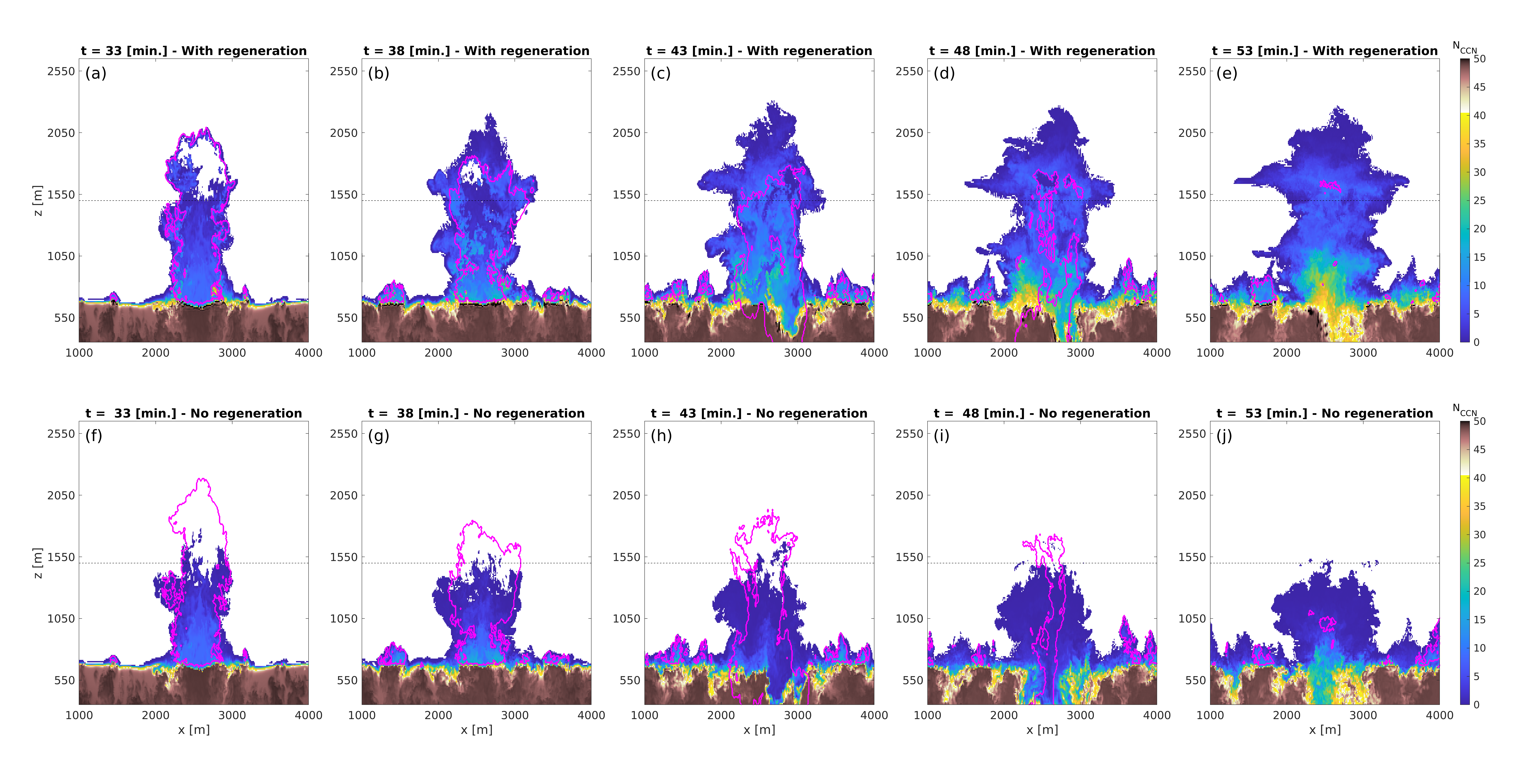}\\
  \caption{Dry aerosol concentration [$\frac{\#}{cm^3}$] for the clean cloud case, at different times.  The magenta contour denotes the cloud boundary (defined by a threshold of  $LWC>0.01 [\frac{g}{kg}]$), and the horizontal black dashed line represents the inversion layer base. The upper (lower) row represents the results of the simulation with (without) the regeneration scheme. The initial dry aerosol number concentration was $50 cm^{-3}$, distributed below $600m$. }
  \label{Nccn50}
 \end{figure}

\begin{figure}[H]

  % \noindent\includegraphics[width=19pc,angle=0]{Nccn_crossction_32to52_500CCN_3.jpg}\\
  \noindent\includegraphics[width=\linewidth]{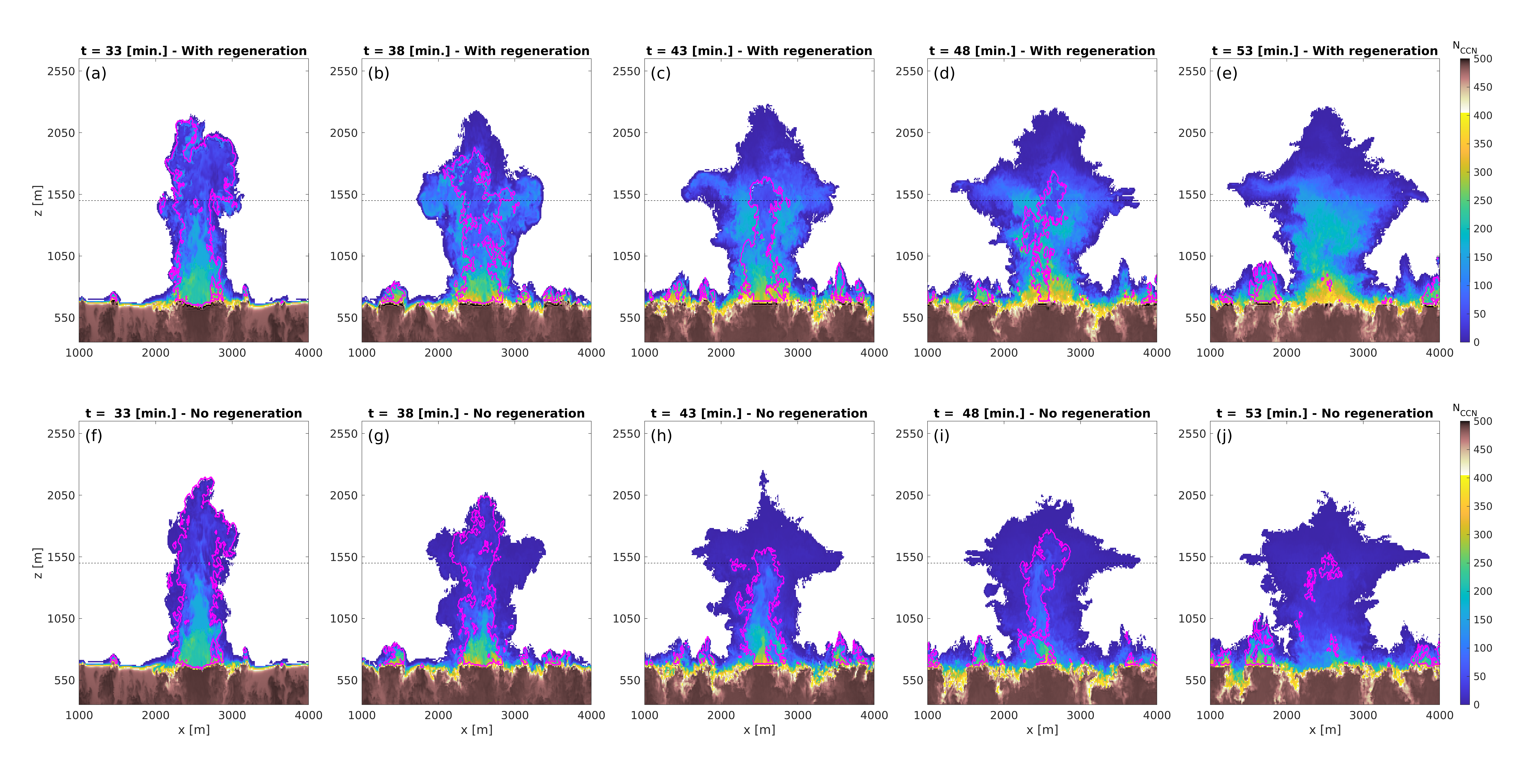}\\
  \caption{Same as Figure \ref{Nccn50} but for the polluted cloud. Note the different color scale. }
  \label{Nccn500}
\end{figure}

Figures \ref{Nccn50} (clean case) and \ref{Nccn500} (polluted case) present the cross-sections of the two simulated clouds and the dry aerosol concentration around them at different times along their dissipation stage (after 33 min; the time of clouds' maximal depth). During the dissipation stage, there is intensive evaporation, and the regeneration of aerosols enriches their concentration around the clouds. 

There is a significant difference in the concentration and altitude of the aerosols between the simulated results with (Figure \ref{Nccn50}, upper row) and without (bottom row) aerosol regeneration in the clean conditions simulations. In the simulation without the regeneration scheme, the aerosols are confined to an altitude of $1500 m$, in contrast to the regeneration simulation, where there are more aerosols and in higher altitudes. The reasons for these differences lie in the microphysical and dynamic cloud processes. In both cases, the big aerosols were activated in the lower part of the cloud, and only smaller ones were carried upward by the updrafts. The supersaturation conditions that developed at elevated levels (above $1500 m$), enabled the activation of small dry aerosols in both cases (with and without aerosol regeneration; Figure \textbf{S1}a,f). As a result, the concentration of non-activated dry aerosols approached zero in the simulation without regeneration. In contrast, dry aerosols are observed at high altitudes in the simulation with aerosol regeneration (Figure \ref{Nccn50}b-e), as they were created through droplet evaporation. Figure \textbf{S3}b,c, shows the presence of downdrafts at high levels, which leads to this droplet evaporation.

In the polluted cloud simulations (Figure \ref{Nccn500}), there is a large concentration of drops, resulting in lower supersaturation values (Figure \textbf{S2}a,f) compared to the clean case. Consequently, there is less activation of the smallest dry aerosols; instead, they are advected upward with the cloud updraft to the cloud-top (lower row of Figure \ref{Nccn500}) to altitudes of approximately $2000 m$, both with and without considering the regeneration process. At the same time, drops evaporation results in a much larger dry aerosol concentration in the regeneration case (upper row of Figure \ref{Nccn500}). It is important to note that the polluted cloud does not precipitate, allowing for a high concentration of dry aerosols, reaching several hundred per cubic centimeter after droplet evaporation.\\

Precipitation is formed in the clean cloud simulations (Figure \ref{Nccn50}). The simulation without regeneration produces a larger amount of surface rain (a total amount of $\sim38.9 \cdot 10^3$ kg) compared to the simulation with regeneration, which produces a total of $\sim18.6 \cdot 10^3$ kg. This difference can be explained by the entrainment of dry aerosols into the cloud in the regeneration simulation. This increases the droplet concentration and intensifies the competition for available water vapor, resulting in smaller droplets. This reduces the collision coalescence efficiency, ultimately translating into a reduced rainfall amount.
In the clean cloud case, the reduced dry aerosol amount below the cloud base ($600 m$) can be attributed to the strong downdrafts caused by the rain (Figure \ref{Nccn50}c-e, h-j, and Figure \textbf{S3}c,h).

The evolution of all clouds simulated in this work can be described, in general, in a similar way. The growing stage lasts up to approximately 33 min when the clouds reach their maximum depth, followed by dissipation. While the top of the growing cloud is moving inside the inversion layer ($1500-2000 m$), the buoyancy in its upper part is negative (\cite{ShallowCumulusPropertiesasCapturedbyAdiabaticFractioninHighResolutionLESSimulations}), causing development of in-cloud downdrafts (see in Figure \textbf{S3}b,g, and Figure \textbf{S4}b,g). In parallel, there is a strong detrainment at the upper part of the cloud. The detrainment supports drop evaporation, which returns dry aerosols to the environment. This detrainment zone can be seen both in Figure \ref{Nccn50} and \ref{Nccn500} in panels c-e in the anvil-like shape of the dry aerosol concentration. The most intense droplet evaporation occurs in the inversion zone, where vigorous cloud-environment mixing results in cloud dilution and sub-saturation conditions.\\ 

%[maybe can delete this part as well]
%There is a substantial difference in the vertical velocity fields between clean-precipitating (Figure \textbf{S3}) and polluted-non-precipitating (Figure \textbf{S4}) clouds. In the clean clouds case, a negative vertical velocity dominates at 38-43 min. The precipitation intensifies the downdrafts (Figure \textbf{S3}c,h). In contrast, in the polluted clouds, the downdrafts are caused by a negative buoyancy (when the cloud penetrates the dry inversion layer) and by evaporative cooling (Figure \textbf{S4}b,g).\\

%The vertical velocity field, as depicted in Figure \ref{w500}d,e, reveals the presence of gravity waves generated by the dissipating cloud. These gravity waves result from the interaction of the decaying cloud with the surrounding atmosphere and propagate through the stable, stratified atmosphere. Further investigation into this subject is warranted to deepen our understanding.

\begin{figure}[H]
  % \noindent\includegraphics[width=19pc,angle=0]{Nccn_1x1_max_avg_32to52_50CCN.jpg}\\
  \noindent\includegraphics[width=\linewidth]{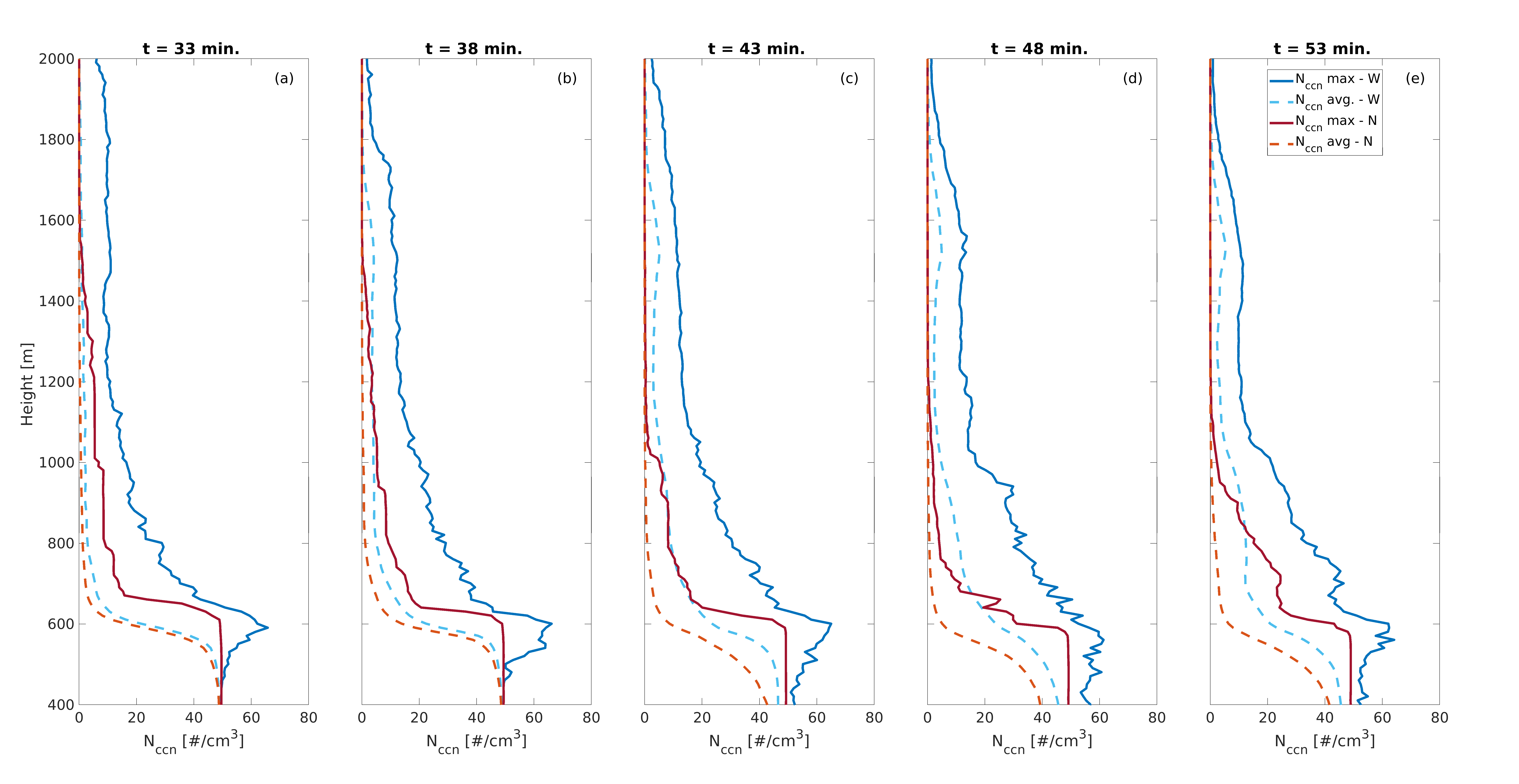}\\
  \caption{Vertical profiles of the maximum and average dry aerosol concentration in the clean cloud case at different times. The dark blue (red) lines represent the vertical profile of the maximal $N_{CCN}$ for the simulation with (without) the aerosol regeneration scheme. The dashed light blue (orange) lines represent the average $N_{CCN}$ profile for each level for the simulation with (without) the aerosol regeneration. Taken for an area of $1$  $km^2$  around the cloud's center.}
  \label{Nccn_horiz50}
\end{figure}

\begin{figure}[H]
  % \noindent\includegraphics[width=19pc,angle=0]{Nccn_1x1_max_avg_32to52_500CCN.jpg}\\
  \noindent\includegraphics[width=\linewidth]{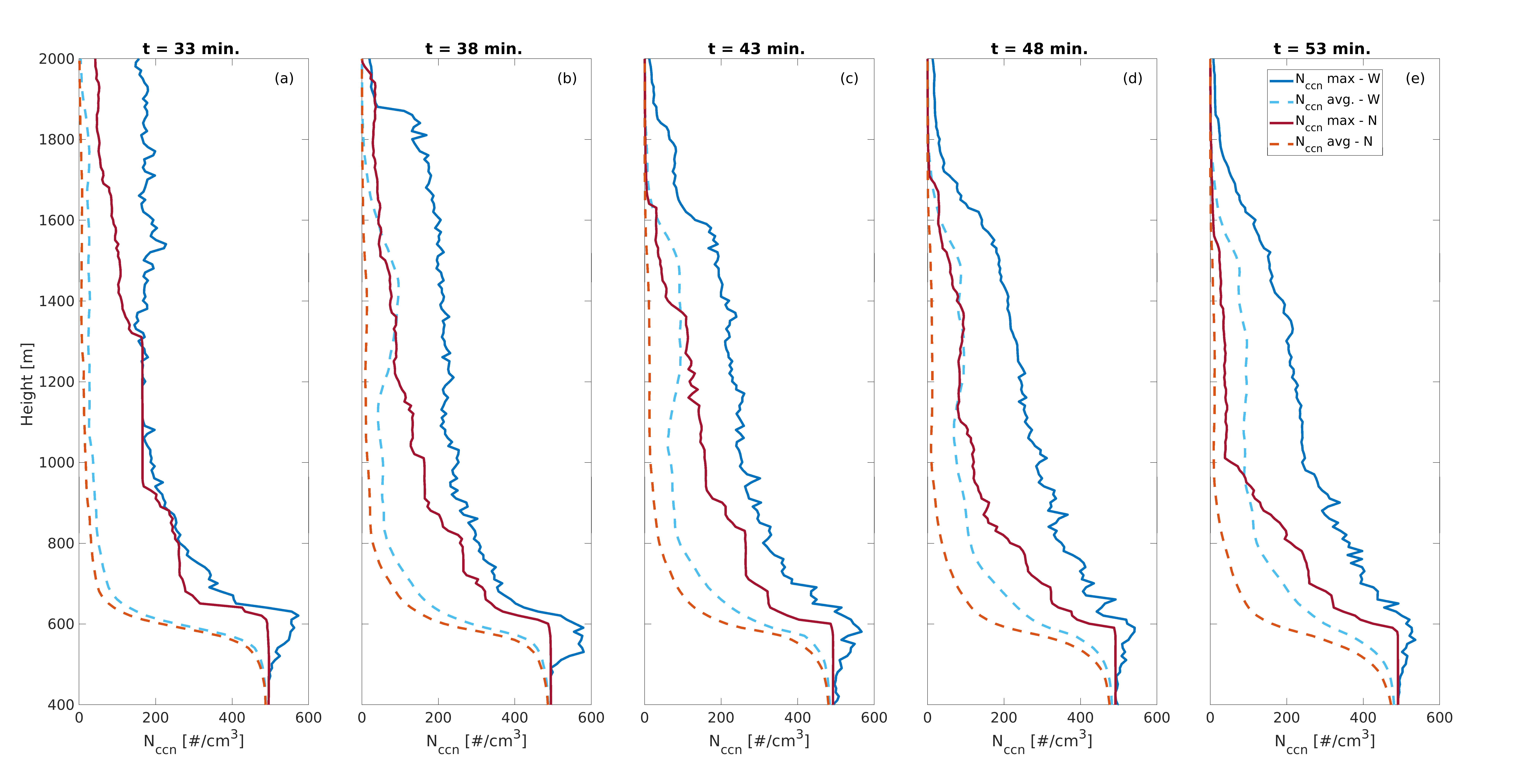}\\
  \caption{Same as Figure \ref{Nccn_horiz50}, for the polluted cloud case. Notice the different scales of the axes compared to Figure \ref{Nccn_horiz50}.}
  \label{Nccn_horiz500}
\end{figure}

Figures \ref{Nccn_horiz50} and \ref{Nccn_horiz500} present the vertical profiles of the mean and maximum (per specific altitude) dry aerosol number concentration ($N_{CCN}$) at different times, along the dissipation stage in the clean (Figure \ref{Nccn_horiz50}) and polluted (Figure \ref{Nccn_horiz500}) clouds cases. Both the maximum and the mean values of $N_{CCN}$ are larger in the regeneration simulations. As dissipation progresses, the curve depicting the mean $N_{CCN}$ values for the regeneration simulation (dashed light blue line) diverges further from the curve for the simulation without regeneration (dashed orange line). This indicates a significant return of dry aerosols to the atmosphere following droplet evaporation. \\

In the regeneration simulations, the values of maximum $N_{CCN}$ below the cloud base exceed the initial ones due to the addition of aerosols released by drop evaporation on top of the background values. \\

In summary, the simulations that treated the aerosol regeneration showed a significantly larger dry aerosol concentration compared to the simulation that didn't consider it. This is especially true for non-precipitating clouds. These results demonstrate the significant impact of cloud processes on the vertical profiles of aerosol concentration in the convective boundary layer.

%This characteristic vertical scale is close to the 1-2 km scale of continental aerosols, which was found in measurements (\cite{jaenicke1993tropospheric}; \cite{khain_book}, fig. 2.2.9).These results demonstrate the significant impact of cloud processes on the vertical profiles of aerosol concentration in the convective boundary layer.

\subsection{Effect of Clouds on Aerosol Size}
\begin{figure}[H]
  % \noindent\includegraphics[width=19pc,angle=0]{fccn_48min_50CCN.jpg}\\
  % \noindent\includegraphics[width=\linewidth]{Fccn_50.jpg}\\
  \noindent\includegraphics[width=\linewidth]{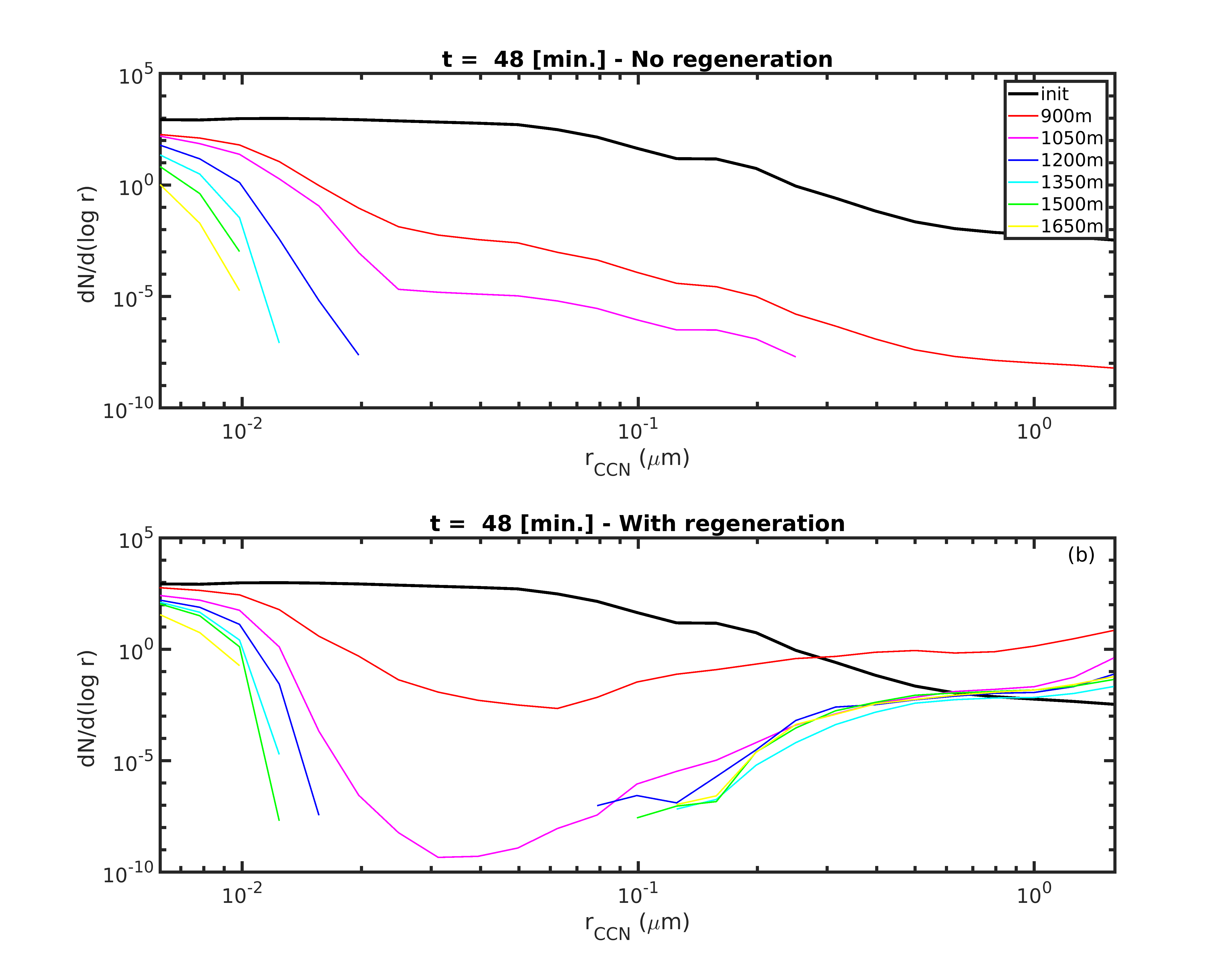}\\
  \caption{Dry aerosol size distribution [$\frac{\#}{cm^3}$], at different heights, for the clean cloud, at 48 min of simulation. Each line represents the mean of 30 size distributions taken from randomly chosen 30 places at a certain height, outside the cloud, up to $\sim1000 [m]$ around the cloud center. The black line denotes the initial dry aerosol size distribution below the cloud base. The upper (bottom) panel represents the simulation without (with) aerosol regeneration. }
  \label{fccn50}
\end{figure}

\begin{figure}[H]
  % \noindent\includegraphics[width=19pc,angle=0]{fccn_48min_500CCN.jpg}\\
  \noindent\includegraphics[width=\linewidth]{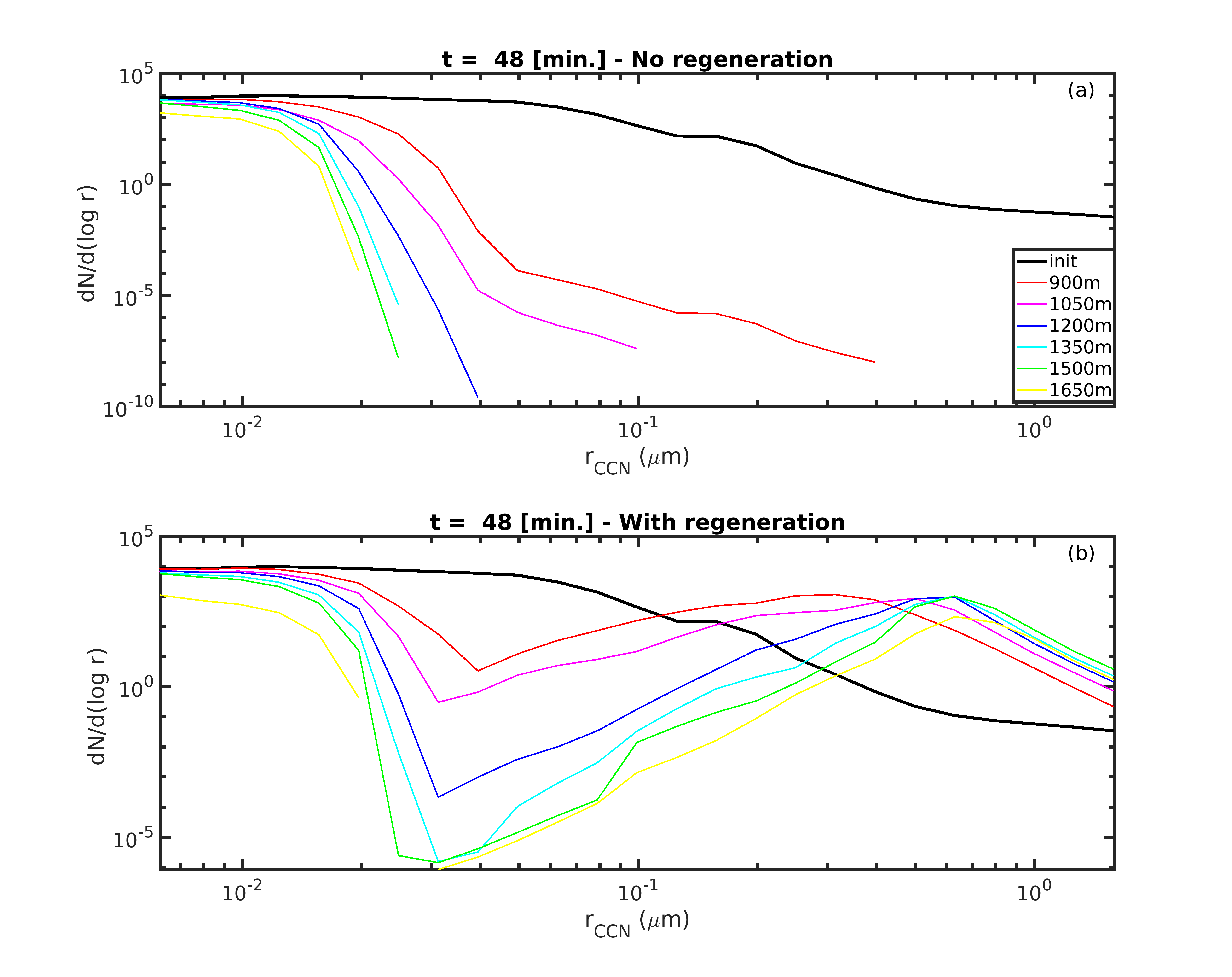}\\
  \caption{Same as Figure \ref{fccn50} but for a polluted cloud. }
  \label{fccn500}
\end{figure}

Next, we will examine the effect of the regeneration process on the aerosol size distributions. Figures \ref{fccn50} (clean cloud) and \ref{fccn500} (polluted cloud) present the size distribution of the dry aerosols at different heights at 48 min of simulation (advanced dissipation stage). Figures \ref{fccn50}\textbf{a} and  \ref{fccn500}\textbf{a} show the results of the simulations that did not consider the regeneration scheme; hence, they present the dry aerosols that were not activated and were advected with the air. Figures \ref{fccn50}\textbf{b} and \ref{fccn500}\textbf{b} present the simulated results with the regeneration scheme. These panels show a significant enhancement in the concentration of large aerosols compared to the size distribution in the initial stage. These large aerosols formed through the collision-coalescence of droplets that eventually evaporated. This is part of the significant effect of the regeneration process on aerosol size distribution that can be easily recognized. The regeneration process leads to a bimodal aerosol spectrum. The first mode consists of the smallest non-activated aerosols, which ascend within the cloud's updrafts. The width of this mode decreases with height, which indicates in-cloud nucleation, meaning the transformation of the largest dry aerosols into drops. The second mode results from the regeneration of aerosols through droplet evaporation, which occurs after collisions and coalescence. \\

It can be seen that the first mode of the polluted cloud (Figure \ref{fccn500}) is wider than the one of the clean cloud (Figure \ref{fccn50}). This is due to the larger supersaturation values in the clean cloud case that caused the activation of more small-mode aerosols into droplets. \\ 

In the clean simulation case (Figure \ref{fccn50}\textbf{b}), the concentration of large dry aerosols increases with their size, and there are no median-size aerosols. This is a result of the collision-coalescence of drops. In addition, the concentration of the large dry aerosols decreases with height in the clean clouds case since they efficiently precipitate, and most of the raindrops fall down in the cloud (they don't evaporate). The picture changes substantially when examining the dry aerosol within and around the polluted cloud (Figure \ref{fccn500}\textbf{b}). Firstly, the second mode in the size distribution has a maximum at $0.4-0.7 \mu m$, and then the number decreases for larger sizes. This is due to the small sizes of drops in the polluted case that collide and coalesce, forming smaller drops compared to the drops resulting from collisions in the clear cloud case. Secondly, the concentration of large dry aerosols tends to increase with height. This can be explained by more collisions as a function of height in the cloud, leading to the regeneration of larger dry aerosols in case of evaporation, with no rain to deplete them. \\

Observation of two modes in the dry aerosol size distribution was previously reported from in-situ measurements and referred to as the 'Hoppel minima' (\cite{hoppel1986effect,hoppel1990submicron}).  This minimum in the dry aerosol size distribution between the two modes is closely related to the bimodal distribution of cloud droplets and raindrops and attributed to collisions between them.

\begin{figure}[H]
  % \noindent\includegraphics[width=19pc,angle=0]{Nccn_horizontal_crossction_48min_1500m_50CCN_4.jpg}\\
  \noindent\includegraphics[width=\linewidth]{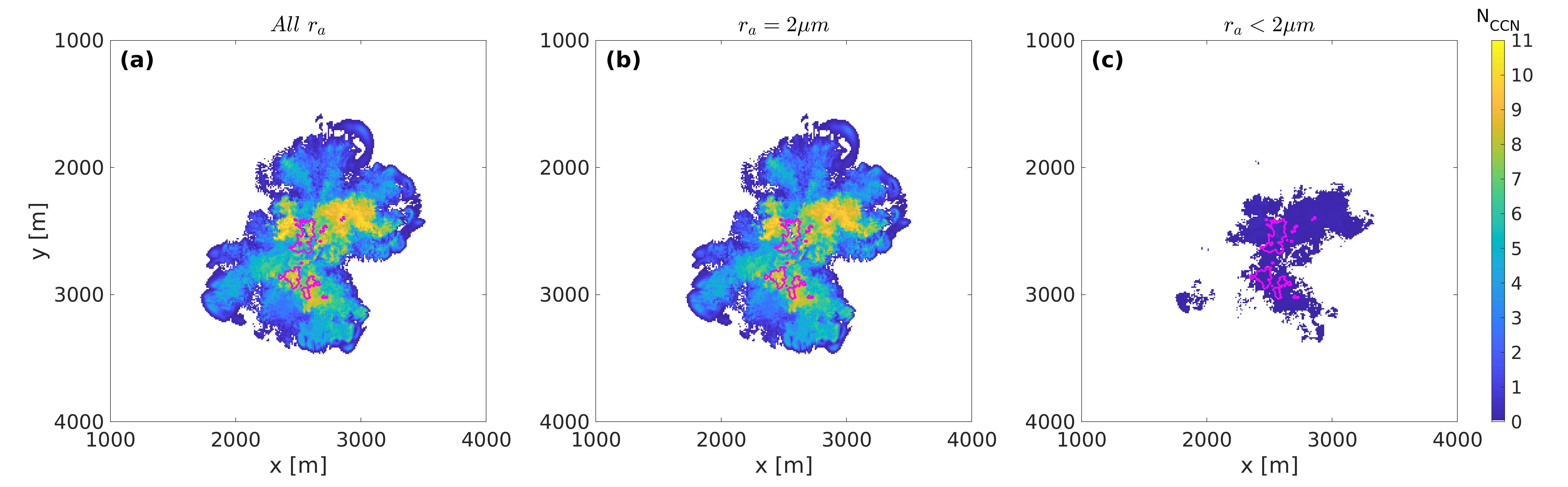}\\
  \caption{Horizontal cross-sections of dry aerosol concentration [$\frac{\#}{cm^3}$], at $1500 m$ height and $48 min$ of simulation for the clean cloud case, from the simulation with the regeneration of aerosols. The magenta contour marks the cloud boundary.  Panel \textbf{a} is the total dry aerosol concentration, including all bins. Panel \textbf{b} includes only the bins with a radius equal $2 \mu m$ (the biggest bin size in the grid),  Panel \textbf{c} includes only bins with a smaller radius than $2 \mu m$.}
  \label{horiNccn50}
\end{figure}

\begin{figure}[H]
  % \noindent\includegraphics[width=19pc,angle=0]{Nccn_horizontal_crossction_48min_1500m_500CCN_3.jpg}\\
  \noindent\includegraphics[width=\linewidth]{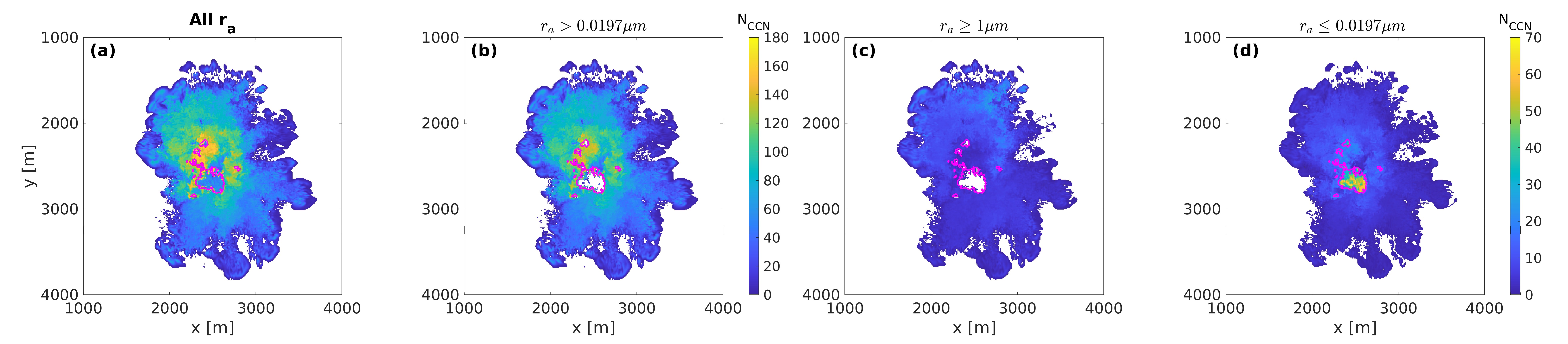}\\
  \caption{Horizontal cross-sections of dry aerosol concentration [$\frac{\#}{cm^3}$], at $1500 m$ height and $48 min$ of simulation for the polluted cloud case, from the simulation with the regeneration of aerosols. The magenta contour marks the cloud boundary. Panel \textbf{a} is the total dry aerosol concentration, including all bins. Panel \textbf{b} includes only bins with a radius bigger than $0.0197 \mu m$,  Panel \textbf{c} includes only bins with a radius equal or bigger than $1 \mu m$, and panel \textbf{d} includes only bins with a radius equal or smaller than $0.0197 \mu m$. The value of $r_a=0.0197 \mu m$ was chosen because it is the largest dry aerosol that was not activated at that snapshot and at that height. }
  \label{horiNccn500} 
\end{figure}

Figures \ref{horiNccn50} and \ref{horiNccn500} present horizontal cross-sections (at $1500 m$) of the dry aerosol number concentration for the clean and polluted clouds. They correspond to a simulation time of 48 min, similar time to the results presented in the previous Figures (Figure \ref{fccn50} and \ref{fccn500}).

Note that the simulated cumulus cloud creates a large zone of enhanced dry aerosol concentration around it, with approximate linear dimensions of $\sim2 km$ in the clean case (Figure \ref{horiNccn50}) and $\sim2.5 km$ in the polluted case (Figure \ref{horiNccn500}).  These regions around the clouds are 2-3 times larger than the clouds at their maximal size (t$\sim33$ min). Accordingly, the clouds generate an aerosol volume 4-9 times larger than that of mature clouds. This large aerosol volume is created by toroidal circulations that determine the detrainment and entrainment in clouds. Those circulations are particularly pronounced within the inversion layer (\cite{khain2024dynamics,arieli2024distinct}).

In the clean case simulation (Figure \ref{horiNccn50}a,b), most of the dry aerosols around the cloud have a radius of $r_a = 2\mu m$. Note that in our model, $r_a = 2\mu m$ is the maximum dry aerosol bin size; hence, the actual size of the aerosol might have been even larger (due to drop collisions). Contrary, in the polluted case (Figure \ref{horiNccn500}c), only a small number of dry aerosols (out of the total) with $r_a\geq1\mu m$ were detected in this height. 
Figure \ref{horiNccn500}b,c shows that inside the cloud (marked by a magenta counter), there are only small dry aerosols, with $r_a<0.0197\mu m$. These small dry aerosols were not activated and were carried to this height by the cloud's updraft. This agrees with the discussion in subsection \textbf{a}.

While the simulations were initialized by locating the dry aerosols only below the cloud base (up to $600m$) and mainly within the accumulation mode size ($0.006-0.5 \mu m$ aerosol radius size), at the dissipation stage, the aerosols are situated up to $\sim1800 m$, with a shift of the size distribution toward larger sizes. This shift due to the collision-coalescence process agrees with previous studies (\cite{flossmann1985theoretical,feingold1996numerical,hoffmann2023note}).\\ 

\begin{figure}[H]
  % \noindent\includegraphics[width=19pc,angle=0]{Nccn_horizontal_crossction_48min_1500m_500CCN_3.jpg}\\
  \noindent\includegraphics[width=\linewidth]{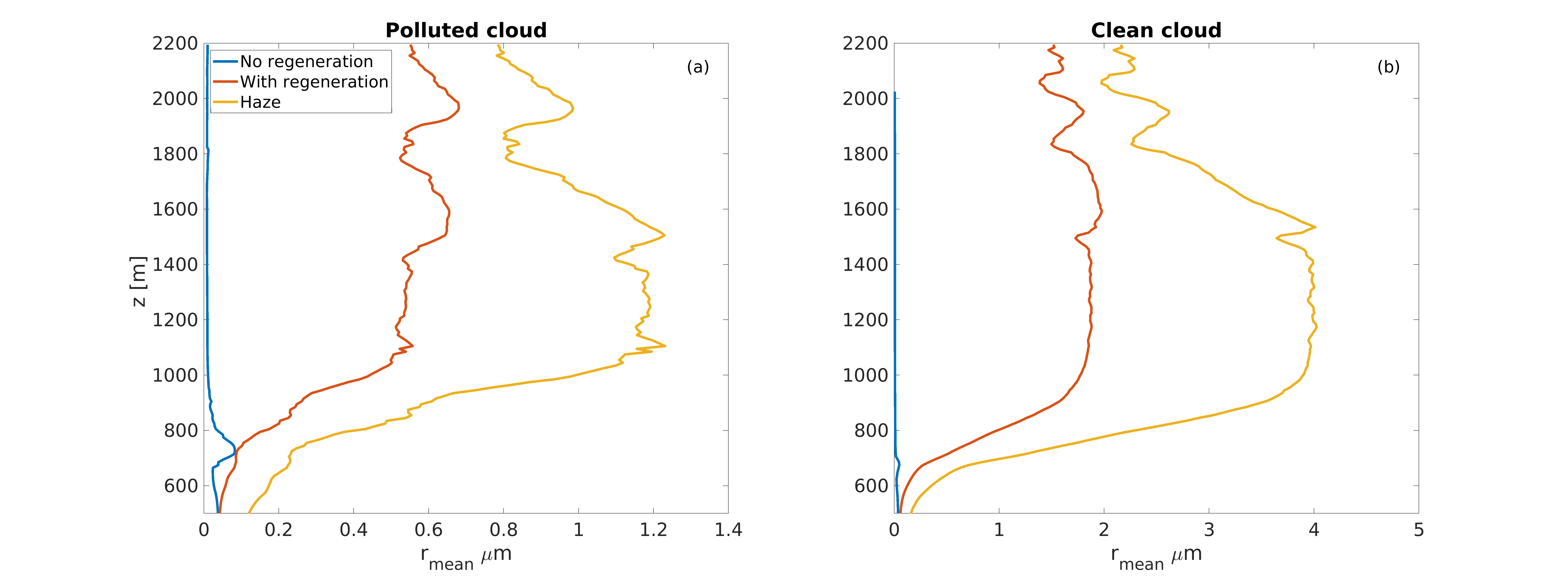}\\
  \caption{Vertical profiles of the mean dry aerosol radius and mean haze radius (per height) at 52 min of simulation. The blue (red) line is the mean dry aerosol radius for the simulation without (with) the regeneration of aerosols. The yellow line represents the mean haze radius in the simulation with the regeneration. Panel \textbf{a} is for the polluted cloud, and panel \textbf{b} is for the clean cloud case. The calculation of the mean includes the voxels in sub-saturation conditions in a square of $2$ $km^2$  centered around the cloud center.}
  \label{rmean}
\end{figure}

As shown in Figures \textbf{S1} and \textbf{S2}, the drop evaporation increases the relative humidity in the environment surrounding the cloud. These humidity levels exceed the delinquency level,  i.e., soluble aerosol particles produce haze with a size that can be calculated using eq. (1).
Figure \ref{rmean} presents the vertical profiles of the mean size (per height) of the dry aerosol and of the haze at 52 min of simulation (panel \textbf{a} - polluted, panel \textbf{b} - clean). For both cases, the profile of the mean radius in the simulations that included the regeneration grows exponentially from the cloud base to about $1000 m$ height. It presents a similar mean radius at higher altitudes. In the polluted case, the maximum mean dry radius is $\sim0.6 \mu m$ (at $1500-1700 m$) while the maximum mean haze radius is $\sim1.2 \mu m$ (at $1100-1500 m$), in the clean case, the maximum mean dry radius is $\sim1.9 \mu m$ (around $1600 m$) while the maximum mean haze radius is $\sim4 \mu m$ (at $1100-1500 m$). 

Our findings show a much larger mean dry aerosol radius for the simulations that considered the regeneration process. This occurs as some of the big activated aerosols and the merged aerosols after drop collisions return to the free atmosphere after drop evaporation. Notably, the haze size is larger than that of the dry aerosol, highlighting the regeneration mechanism's crucial role in influencing the troposphere's radiation properties.

\section{Conclusions}
In this study, we used the \textbf{S}ystem for \textbf{A}tmospheric \textbf{M}odeling (SAM), coupled with the Spectral Bin Microphysical (SBM). The detailed representation of the microphysical process by the SBM model,  along with the high-resolution simulations that capture important scales of turbulent flow, allows for an accurate representation of the life cycle of aerosol particles within cloud droplets. To track the aerosols and their changes due to the microphysical processes experienced by the cloud droplets in which they are embedded, we added another size distribution grid to the model: the mean dry aerosol mass per droplet size bin.  This enables us to investigate the effects of clouds on aerosol concentrations and properties. We simulate shallow cumulus clouds under different aerosol conditions (initialized below cloud base): $50 cm^{-3}$ (clean - precipitating) and $500 cm^{-3}$ (polluted - non-precipitating). For each type of cloud, two simulations were conducted, with and without the regeneration scheme that brings aerosols back to the atmosphere after drop evaporation. This study examines the effects of clouds on atmospheric aerosols, specifically focusing on the cloud dissipation stage since drop evaporation and aerosol regeneration occur mainly at this stage in the cloud's lifecycle. \\  

Variations in dry aerosol concentrations and sizes impact cloud processes and properties, potentially leading to changes in cloud size (including cloud top height), radiative properties, and precipitation processes. These changes can impact cloud processes and evolution through entrainment processes, as well as the formation and processes of consequent clouds within the field, which will meet modified environmental conditions. The interplay between dry aerosol concentration in the boundary layer and cloud microphysics warrants investigation to improve our understanding of boundary layer-aerosol-cloud interactions.\\

Our findings demonstrate significant differences in aerosol vertical distribution and size distribution between the simulations with and without the regeneration process. We found a significant concentration of dry aerosols within the inversion layer of both precipitating and non-precipitating clouds, attributed to aerosol regeneration resulting from intense evaporation. One key finding is the production of giant CCNs by aerosol regeneration after collision-coalescence of drops within the cloud. Notably, we observed the presence of these giant CCNs at altitudes extending up to the inversion layer. These giant CCNs can later re-enter the cloud, influencing its dynamic and microphysical properties. Previous studies have demonstrated the impact of giant CCNs on precipitation processes and on the radiative properties of the atmosphere  \cite{yin2000effects,cheng2007modelling,cheng2009influence} \\

Our results are supported by observational evidence regarding the cloud's impact on aerosol concentration and size. For instance, \cite{dipu2013impact} identified a significant concentration of large aerosols at high altitudes within specific atmospheric layers, namely at 2 and 5 km, which we believe to be inversion layers. The presence of giant CCNs and haze at altitudes of a few kilometers was also observed in the Cloud Aerosol Interaction and Precipitation Enhancement Experiment (Prabhakaran, 2023 personal communication).

Additionally, our study showed a bimodal distribution of the dry aerosol when considering the regeneration process. The phenomenon of two modes with minima in the aerosol size distribution, as depicted in Figures \ref{fccn50} and \ref{fccn500}, is consistent with previous in-situ measurements, known as the 'Hoppel minima' (\cite{hoppel1986effect,hoppel1990submicron,liu2021sea}). This phenomenon is attributed to collision-coalescence of cloud droplets, a mechanism supported by both theoretical   (\cite{flossmann1985theoretical}) and numerical (\cite{hoffmann2023note}) studies.\\

The results of our study highlight the profound influence that even a single cumulus cloud can have on the concentration of dry aerosols in the boundary layer. We demonstrated that a cloud with a scale of approximately 1 km creates a surrounding region with altered aerosol concentrations that extend 2-3 times farther than the cloud's horizontal extent. The components comprising the cloud twilight zone are typically considered a combination of humidified aerosol (\cite{bar2012radiative,twohy2009effect}), undetected clouds (\cite{eytan2020longwave,hirsch2014transition}), and larger processed aerosol (\cite{marshak2021aerosol,eck2012fog}). However, our findings suggest an additional contributing factor: increased aerosol concentrations around clouds due to aerosol transport facilitated by the clouds themselves, thereby substantially shaping the Earth's radiative properties.\\ 

%This effect can be particularly pronounced in the twilight zone, substantially shaping the Earth's radiative properties.\\

%[[I think you don't need the next paragraph]]
%It's worth noting that aerosols are transported upward by turbulence, large-scale updrafts, or dust storms, which may not be directly associated with clouds (\cite{dipu2013impact}). These aerosols vary in size, ranging from small to large. Therefore, this study only presents a portion, though it is an important one, of the complete picture regarding aerosol vertical distributions in the atmosphere. \\

%Recent works reported that a substantial fraction of droplets originates by activation of lateral entrained aerosols (\cite{chandrakar2021impact,hoffmann2015entrainment}. 

In conclusion, our study sheds more light on the complex relationship between aerosol dynamics, concentration, and cloud processes. It highlights the importance of integrating these factors into atmospheric modeling and emphasizes the significant role of dry aerosol regeneration in cloud microphysics and vertical aerosol transport. Understanding the changes in the vertical profile of aerosol and its size distribution is vital for improving our predictions of cloud processes and properties in various environmental contexts. Further research in this area is essential for advancing our understanding of cloud-aerosol interactions and their broader implications for weather and climate. The subsequent stage of our research will focus on examining how the regeneration of aerosols affects the cloud microphysical processes.

\clearpage
%%%%%%%%%%%%%%%%%%%%%%%%%%%%%%%%%%%%%%%%%%%%%%%%%%%%%%%%%%%%%%%%%%%%%
% ACKNOWLEDGMENTS
%%%%%%%%%%%%%%%%%%%%%%%%%%%%%%%%%%%%%%%%%%%%%%%%%%%%%%%%%%%%%%%%%%%%%
\acknowledgments
This project has received funding from the European Research Council (ERC) under the European Union’s Horizon 2020 research and innovation programme (CloudCT, grant agreement No 810370), and by The Israel Science Foundation (grants 2635/20; 1449/22).
%  Keep acknowledgments (note correct spelling: no ``e'' between the ``g'' and
% ``m'') as brief as possible. In general, acknowledge only direct help in
%  writing or research. Financial support (e.g., grant numbers) for the work done, 
%  for an author, or for the laboratory where the work was performed must be 
%  acknowledged here rather than as footnotes to the title or to an author's name.
%  Contribution numbers (if the work has been published by the author's institution 
%  or organization) should be placed in the acknowledgments rather than as 
%  footnotes to the title or to an author's name.

%%%%%%%%%%%%%%%%%%%%%%%%%%%%%%%%%%%%%%%%%%%%%%%%%%%%%%%%%%%%%%%%%%%%%
% DATA AVAILABILITY STATEMENT
%%%%%%%%%%%%%%%%%%%%%%%%%%%%%%%%%%%%%%%%%%%%%%%%%%%%%%%%%%%%%%%%%%%%%
% 
%
\datastatement
The SAM codes are available on the website of Prof. Marat Khairoutdinov \cite{SAMref}.
The codes to reproduce the figures of the manuscript are available are publicly available at \url{https://doi.org/10.34933/871eae61-de6f-4274-9515-b829611e0141}.
%  The data availability statement is where authors should describe how the data underlying 
%  the findings within the article can be accessed and reused. Authors should attempt to 
%  provide unrestricted access to all data and materials underlying reported findings. 
%  If data access is restricted, authors must mention this in the statement. See
%  {http://www.ametsoc.org/PubsDataPolicy} for more info.

%%%%%%%%%%%%%%%%%%%%%%%%%%%%%%%%%%%%%%%%%%%%%%%%%%%%%%%%%%%%%%%%%%%%%
% APPENDIXES
%%%%%%%%%%%%%%%%%%%%%%%%%%%%%%%%%%%%%%%%%%%%%%%%%%%%%%%%%%%%%%%%%%%%%
%
%% If only one appendix, use

%\appendix

%% If more than one appendix, use \appendix[<letter>], e.g.,

%\appendix[A] 

%% Appendix title is necessary! For appendix title:

%\appendixtitle{Title of Appendix}

%%% Appendix section numbering (note, skip \section and begin with \subsection)
%
% \subsection{First primary heading}

% \subsubsection{First secondary heading}

% \paragraph{First tertiary heading}

%%%%%%%%%%%%%%%%%%%%%%%%%%%%%%%%%%%%%%%%%%%%%%%%%%%%%%%%%%%%%%%%%%%%%
% REFERENCES
%%%%%%%%%%%%%%%%%%%%%%%%%%%%%%%%%%%%%%%%%%%%%%%%%%%%%%%%%%%%%%%%%%%%%
% Make your BibTeX bibliography by using these commands:
 \bibliographystyle{unsrt}
 \bibliography{bibyael}

\begin{thebibliography}{58}
\providecommand{\natexlab}[1]{#1}
\providecommand{\url}[1]{\texttt{#1}}
\renewcommand{\UrlFont}{\rmfamily}
\providecommand{\urlprefix}{URL }
\expandafter\ifx\csname urlstyle\endcsname\relax
  \providecommand{\doi}[1]{https://doi.org/\discretionary{}{}{}#1}\else
  \providecommand{\doi}{https://doi.org/\discretionary{}{}{}\begingroup \urlstyle{rm}\Url}\fi
\providecommand{\eprint}[2][]{\url{#2}}

\bibitem[{Albrecht(1989)}]{albrecht1989aerosols}
Albrecht, B.~A., 1989: Aerosols, cloud microphysics, and fractional cloudiness. \textit{Science}, \textbf{245~(4923)}, 1227--1230.

\bibitem[{Altaratz et~al.(2008)Altaratz, Koren, Reisin, Kostinski, Feingold, Levin,, and Yin}]{altaratz2008aerosols}
Altaratz, O., I.~Koren, T.~Reisin, A.~Kostinski, G.~Feingold, Z.~Levin, and Y.~Yin, 2008: Aerosols' influence on the interplay between condensation, evaporation and rain in warm cumulus cloud. \textit{Atmospheric Chemistry and Physics}, \textbf{8~(1)}, 15--24.

\bibitem[{Altaratz et~al.(2014)Altaratz, Koren, Remer,, and Hirsch}]{altaratz2014cloud}
Altaratz, O., I.~Koren, L.~Remer, and E.~Hirsch, 2014: Cloud invigoration by aerosols—coupling between microphysics and dynamics. \textit{Atmospheric Research}, \textbf{140}, 38--60.

\bibitem[{Arieli et~al.(2024)Arieli, Eytan, Altaratz, Khain,, and Koren}]{arieli2024distinct}
Arieli, Y., E.~Eytan, O.~Altaratz, A.~Khain, and I.~Koren, 2024: Distinct mixing regimes in shallow cumulus clouds. \textit{Geophysical Research Letters}, \textbf{51~(2)}, e2023GL105\,746.

\bibitem[{Bar-Or et~al.(2012)Bar-Or, Koren, Altaratz,, and Fredj}]{bar2012radiative}
Bar-Or, R., I.~Koren, O.~Altaratz, and E.~Fredj, 2012: Radiative properties of humidified aerosols in cloudy environment. \textit{Atmospheric research}, \textbf{118}, 280--294.

\bibitem[{Beard(1976)}]{beard1976terminal}
Beard, K.~V., 1976: Terminal velocity and shape of cloud and precipitation drops aloft. \textit{Journal of the Atmospheric Sciences}, \textbf{33~(5)}, 851--864.

\bibitem[{Benmoshe et~al.(2012)Benmoshe, Khain, Pinsky,, and Pokrovsky}]{benmoshe2012turbulent}
Benmoshe, N., A.~Khain, M.~Pinsky, and A.~Pokrovsky, 2012: Turbulent effects on cloud microstructure and precipitation of deep convective clouds as seen from simulations with a 2-d spectral microphysics cloud model. \textit{J. Geophys. Res}, \textbf{117}, D06\,220.

\bibitem[{Bott(1998)}]{bott1998flux}
Bott, A., 1998: A flux method for the numerical solution of the stochastic collection equation. \textit{Journal of the atmospheric sciences}, \textbf{55~(13)}, 2284--2293.

\bibitem[{Bretherton et~al.(2007)Bretherton, Blossey,, and Uchida}]{bretherton2007cloud}
Bretherton, C., P.~N. Blossey, and J.~Uchida, 2007: Cloud droplet sedimentation, entrainment efficiency, and subtropical stratocumulus albedo. \textit{Geophysical research letters}, \textbf{34~(3)}.

\bibitem[{Cess(1975)}]{cess1975global}
Cess, R.~D., 1975: Global climate change: An investigation of atmospheric feedback mechanisms. \textit{Tellus}, \textbf{27~(3)}, 193--198.

\bibitem[{Cheng et~al.(2007)Cheng, Wang,, and Chen}]{cheng2007modelling}
Cheng, C.-T., W.-C. Wang, and J.-P. Chen, 2007: A modelling study of aerosol impacts on cloud microphysics and radiative properties. \textit{Quarterly Journal of the Royal Meteorological Society: A journal of the atmospheric sciences, applied meteorology and physical oceanography}, \textbf{133~(623)}, 283--297.

\bibitem[{Cheng et~al.(2009)Cheng, Carri{\'o}, Cotton,, and Saleeby}]{cheng2009influence}
Cheng, W.~Y., G.~G. Carri{\'o}, W.~R. Cotton, and S.~M. Saleeby, 2009: Influence of cloud condensation and giant cloud condensation nuclei on the development of precipitating trade wind cumuli in a large eddy simulation. \textit{Journal of Geophysical Research: Atmospheres}, \textbf{114~(D8)}.

\bibitem[{Dipu et~al.(2013)Dipu, Prabha, Pandithurai, Dudhia, Pfister, Rajesh,, and Goswami}]{dipu2013impact}
Dipu, S., T.~V. Prabha, G.~Pandithurai, J.~Dudhia, G.~Pfister, K.~Rajesh, and B.~Goswami, 2013: Impact of elevated aerosol layer on the cloud macrophysical properties prior to monsoon onset. \textit{Atmospheric Environment}, \textbf{70}, 454--467.

\bibitem[{Eck et~al.(2012)}]{eck2012fog}
Eck, T.~F., and Coauthors, 2012: Fog-and cloud-induced aerosol modification observed by the aerosol robotic network (aeronet). \textit{Journal of Geophysical Research: Atmospheres}, \textbf{117~(D7)}.

\bibitem[{Eytan et~al.(2022)Eytan, Khain, Pinsky, Altaratz, Shpund,, and Koren}]{ShallowCumulusPropertiesasCapturedbyAdiabaticFractioninHighResolutionLESSimulations}
Eytan, E., A.~Khain, M.~Pinsky, O.~Altaratz, J.~Shpund, and I.~Koren, 2022: Shallow cumulus properties as captured by adiabatic fraction in high-resolution les simulations. \textit{Journal of the Atmospheric Sciences}, \textbf{79~(2)}, 409 -- 428, \doi{10.1175/JAS-D-21-0201.1}, \urlprefix\url{https://journals.ametsoc.org/view/journals/atsc/79/2/JAS-D-21-0201.1.xml}.

\bibitem[{Eytan et~al.(2020)Eytan, Koren, Altaratz, Kostinski,, and Ronen}]{eytan2020longwave}
Eytan, E., I.~Koren, O.~Altaratz, A.~B. Kostinski, and A.~Ronen, 2020: Longwave radiative effect of the cloud twilight zone. \textit{Nature geoscience}, \textbf{13~(10)}, 669--673.

\bibitem[{Fan et~al.(2009)Fan, Ovtchinnikov, Comstock, McFarlane,, and Khain}]{fan2009ice}
Fan, J., M.~Ovtchinnikov, J.~M. Comstock, S.~A. McFarlane, and A.~Khain, 2009: Ice formation in arctic mixed-phase clouds: Insights from a 3-d cloud-resolving model with size-resolved aerosol and cloud microphysics. \textit{Journal of Geophysical Research: Atmospheres}, \textbf{114~(D4)}.

\bibitem[{Feingold et~al.(1996)Feingold, Kreidenweis, Stevens,, and Cotton}]{feingold1996numerical}
Feingold, G., S.~M. Kreidenweis, B.~Stevens, and W.~Cotton, 1996: Numerical simulations of stratocumulus processing of cloud condensation nuclei through collision-coalescence. \textit{Journal of Geophysical Research: Atmospheres}, \textbf{101~(D16)}, 21\,391--21\,402.

\bibitem[{Flossmann et~al.(1985)Flossmann, Hall,, and Pruppacher}]{flossmann1985theoretical}
Flossmann, A.~I., W.~Hall, and H.~Pruppacher, 1985: A theoretical study of the wet removal of atmospheric pollutants. part i: The redistribution of aerosol particles captured through nucleation and impaction scavenging by growing cloud drops. \textit{Journal of Atmospheric Sciences}, \textbf{42~(6)}, 583--606.

\bibitem[{Hartmann et~al.(1992)Hartmann, Ockert-Bell,, and Michelsen}]{hartmann1992effect}
Hartmann, D.~L., M.~E. Ockert-Bell, and M.~L. Michelsen, 1992: The effect of cloud type on earth's energy balance: Global analysis. \textit{Journal of Climate}, \textbf{5~(11)}, 1281--1304.

\bibitem[{Hirsch et~al.(2014)Hirsch, Koren, Levin, Altaratz,, and Agassi}]{hirsch2014transition}
Hirsch, E., I.~Koren, Z.~Levin, O.~Altaratz, and E.~Agassi, 2014: On transition-zone water clouds. \textit{Atmospheric Chemistry and Physics}, \textbf{14~(17)}, 9001--9012.

\bibitem[{Hoffmann and Feingold(2023)Hoffmann, and Feingold}]{hoffmann2023note}
Hoffmann, F., and G.~Feingold, 2023: A note on aerosol processing by droplet collision-coalescence. \textit{Geophysical Research Letters}, \textbf{50~(11)}, e2023GL103\,716.

\bibitem[{Hoffmann et~al.(2015)Hoffmann, Raasch,, and Noh}]{hoffmann2015entrainment}
Hoffmann, F., S.~Raasch, and Y.~Noh, 2015: Entrainment of aerosols and their activation in a shallow cumulus cloud studied with a coupled lcm--les approach. \textit{Atmospheric Research}, \textbf{156}, 43--57.

\bibitem[{Hoppel and Frick(1990)Hoppel, and Frick}]{hoppel1990submicron}
Hoppel, W., and G.~Frick, 1990: Submicron aerosol size distributions measured over the tropical and south pacific. \textit{Atmospheric Environment. Part A. General Topics}, \textbf{24~(3)}, 645--659.

\bibitem[{Hoppel et~al.(1986)Hoppel, Frick,, and Larson}]{hoppel1986effect}
Hoppel, W., G.~Frick, and R.~Larson, 1986: Effect of nonprecipitating clouds on the aerosol size distribution in the marine boundary layer. \textit{Geophysical Research Letters}, \textbf{13~(2)}, 125--128.

\bibitem[{Jaenicke(1988)}]{jaenicke1988aerosol}
Jaenicke, R., 1988: Aerosol physics and chemistry. \textit{Zahlenwerte und Funktionen aus Naturwissenschaften und Technik}, \textbf{4}, 391--457.

\bibitem[{Jahani et~al.(2022)Jahani, Andersen, Calb{\'o}, Gonz{\'a}lez,, and Cermak}]{jahani2022longwave}
Jahani, B., H.~Andersen, J.~Calb{\'o}, J.-A. Gonz{\'a}lez, and J.~Cermak, 2022: Longwave radiative effect of the cloud--aerosol transition zone based on ceres observations. \textit{Atmospheric Chemistry and Physics}, \textbf{22~(2)}, 1483--1494.

\bibitem[{Khain(2009)}]{khain2009notes}
Khain, A., 2009: Notes on state-of-the-art investigations of aerosol effects on precipitation: a critical review. \textit{Environmental Research Letters}, \textbf{4~(1)}, 015\,004.

\bibitem[{Khain et~al.(2008)Khain, BenMoshe,, and Pokrovsky}]{khain2008factors}
Khain, A., N.~BenMoshe, and A.~Pokrovsky, 2008: Factors determining the impact of aerosols on surface precipitation from clouds: An attempt at classification. \textit{Journal of the Atmospheric Sciences}, \textbf{65~(6)}, 1721--1748.

\bibitem[{Khain and Pinsky(2018)Khain, and Pinsky}]{khain_book}
Khain, A., and M.~Pinsky, 2018: \textit{Physical processes in clouds and cloud modeling}. Cambridge University Press.

\bibitem[{Khain et~al.(2024)Khain, Pinsky, Eytan, Koren, Altaratz, Arieli,, and Gavze}]{khain2024dynamics}
Khain, A., M.~Pinsky, E.~Eytan, I.~Koren, O.~Altaratz, Y.~Arieli, and E.~Gavze, 2024: Dynamics and microphysics in small developing cumulus clouds. \textit{Atmospheric Research}, 107454.

\bibitem[{Khain et~al.(2004)Khain, Pokrovsky, Pinsky, Seifert,, and Phillips}]{khain2004simulation}
Khain, A., A.~Pokrovsky, M.~Pinsky, A.~Seifert, and V.~Phillips, 2004: Simulation of effects of atmospheric aerosols on deep turbulent convective clouds using a spectral microphysics mixed-phase cumulus cloud model. part i: Model description and possible applications. \textit{Journal of the atmospheric sciences}, \textbf{61~(24)}, 2963--2982.

\bibitem[{Khain et~al.(2019)}]{khain2019parameterization}
Khain, P., and Coauthors, 2019: Parameterization of vertical profiles of governing microphysical parameters of shallow cumulus cloud ensembles using les with bin microphysics. \textit{Journal of the Atmospheric Sciences}, \textbf{76~(2)}, 533--560.

\bibitem[{Khairoutdinov and Randall(2003)Khairoutdinov, and Randall}]{khairoutdinov_2003}
Khairoutdinov, M.~F., and D.~A. Randall, 2003: Cloud resolving modeling of the arm summer 1997 iop: Model formulation, results, uncertainties, and sensitivities. \textit{Journal of Atmospheric Sciences}, \textbf{60~(4)}, 607--625.

\bibitem[{Khairoutdinov(2004)}]{SAMref}
Khairoutdinov, P.~M., 2004: System for atmospheric modeling. \urlprefix\url{http://rossby.msrc.sunysb.edu/~marat/SAM.html}.

\bibitem[{Kogan(1991)}]{kogan1991simulation}
Kogan, Y.~L., 1991: The simulation of a convective cloud in a 3-d model with explicit microphysics. part i: Model description and sensitivity experiments. \textit{Journal of the Atmospheric Sciences}, \textbf{48~(9)}, 1160--1189.

\bibitem[{Koren et~al.(2009)Koren, Feingold, Jiang,, and Altaratz}]{koren2009aerosol}
Koren, I., G.~Feingold, H.~Jiang, and O.~Altaratz, 2009: Aerosol effects on the inter-cloud region of a small cumulus cloud field. \textit{Geophysical Research Letters}, \textbf{36~(14)}.

\bibitem[{Koren et~al.(2007)Koren, Remer, Kaufman, Rudich,, and Martins}]{koren2007twilight}
Koren, I., L.~A. Remer, Y.~J. Kaufman, Y.~Rudich, and J.~V. Martins, 2007: On the twilight zone between clouds and aerosols. \textit{Geophysical research letters}, \textbf{34~(8)}.

\bibitem[{Lebo and Seinfeld(2011)Lebo, and Seinfeld}]{lebo2011continuous}
Lebo, Z., and J.~Seinfeld, 2011: A continuous spectral aerosol-droplet microphysics model. \textit{Atmospheric Chemistry and Physics}, \textbf{11~(23)}, 12\,297--12\,316.

\bibitem[{Liu et~al.(2021)}]{liu2021sea}
Liu, S., and Coauthors, 2021: Sea spray aerosol concentration modulated by sea surface temperature. \textit{Proceedings of the National Academy of Sciences}, \textbf{118~(9)}, e2020583\,118.

\bibitem[{Magaritz et~al.(2010)Magaritz, Pinsky,, and Khain}]{magaritz2010effects}
Magaritz, L., M.~Pinsky, and A.~Khain, 2010: Effects of stratocumulus clouds on aerosols in the maritime boundary layer. \textit{Atmospheric research}, \textbf{97~(4)}, 498--512.

\bibitem[{Marshak et~al.(2021)}]{marshak2021aerosol}
Marshak, A., and Coauthors, 2021: Aerosol properties in cloudy environments from remote sensing observations: A review of the current state of knowledge. \textit{Bulletin of the American Meteorological Society}, \textbf{102~(11)}, E2177--E2197.

\bibitem[{Morcrette et~al.(2009)}]{morcrette2009aerosol}
Morcrette, J.-J., and Coauthors, 2009: Aerosol analysis and forecast in the european centre for medium-range weather forecasts integrated forecast system: Forward modeling. \textit{Journal of Geophysical Research: Atmospheres}, \textbf{114~(D6)}.

\bibitem[{Pinsky et~al.(2001)Pinsky, Khain,, and Shapiro}]{pinsky2001collision}
Pinsky, M., A.~Khain, and M.~Shapiro, 2001: Collision efficiency of drops in a wide range of reynolds numbers: Effects of pressure on spectrum evolution. \textit{Journal of the atmospheric sciences}, \textbf{58~(7)}, 742--764.

\bibitem[{Ramanathan et~al.(1989)Ramanathan, Cess, Harrison, Minnis, Barkstrom, Ahmad,, and Hartmann}]{ramanathan1989cloud}
Ramanathan, V., R.~Cess, E.~Harrison, P.~Minnis, B.~Barkstrom, E.~Ahmad, and D.~Hartmann, 1989: Cloud-radiative forcing and climate: Results from the earth radiation budget experiment. \textit{Science}, \textbf{243~(4887)}, 57--63.

\bibitem[{Ramanathan and Inamdar(2006)Ramanathan, and Inamdar}]{ramanathan2006radiative}
Ramanathan, V., and A.~Inamdar, 2006: The radiative forcing due to clouds and water vapor. \textit{Frontiers of climate modeling}, 119--151.

\bibitem[{Shpund et~al.(2019)}]{shpund2019simulating}
Shpund, J., and Coauthors, 2019: Simulating a mesoscale convective system using wrf with a new spectral bin microphysics: 1: Hail vs graupel. \textit{Journal of Geophysical Research: Atmospheres}, \textbf{124~(24)}, 14\,072--14\,101.

\bibitem[{Siebesma et~al.(2003)}]{ALargeEddySimulationIntercomparisonStudyofShallowCumulusConvection}
Siebesma, A.~P., and Coauthors, 2003: A large eddy simulation intercomparison study of shallow cumulus convection. \textit{Journal of the Atmospheric Sciences}, \textbf{60~(10)}, 1201 -- 1219, \doi{10.1175/1520-0469(2003)60<1201:ALESIS>2.0.CO;2}, \urlprefix\url{https://journals.ametsoc.org/view/journals/atsc/60/10/1520-0469_2003_60_1201_alesis_2.0.co_2.xml}.

\bibitem[{Squires(1958)}]{squires1958microstructure}
Squires, P., 1958: The microstructure and colloidal stability of warm clouds: Part i—the relation between structure and stability. \textit{Tellus}, \textbf{10~(2)}, 256--261.

\bibitem[{Stephens and Webster(1979)Stephens, and Webster}]{SensitivityofRadiativeForcingtoVariableCloudandMoisture}
Stephens, G.~L., and P.~J. Webster, 1979: Sensitivity of radiative forcing to variable cloud and moisture. \textit{Journal of Atmospheric Sciences}, \textbf{36~(8)}, 1542 -- 1556, \doi{10.1175/1520-0469(1979)036<1542:SORFTV>2.0.CO;2}, \urlprefix\url{https://journals.ametsoc.org/view/journals/atsc/36/8/1520-0469_1979_036_1542_sorftv_2_0_co_2.xml}.

\bibitem[{Sun and Lindzen(1993)Sun, and Lindzen}]{sun1993distribution}
Sun, D.-Z., and R.~S. Lindzen, 1993: Distribution of tropical tropospheric water vapor. \textit{Journal of Atmospheric Sciences}, \textbf{50~(12)}, 1643--1660.

\bibitem[{Tang and Munkelwitz(1994)Tang, and Munkelwitz}]{tang1994aerosol}
Tang, I., and H.~Munkelwitz, 1994: Aerosol phase transformation and growth in the atmosphere. \textit{Journal of Applied Meteorology and Climatology}, \textbf{33~(7)}, 791--796.

\bibitem[{Tang(1997)}]{tang1997thermodynamic}
Tang, I.~N., 1997: Thermodynamic and optical properties of mixed-salt aerosols of atmospheric importance. \textit{Journal of Geophysical Research: Atmospheres}, \textbf{102~(D2)}, 1883--1893.

\bibitem[{Twohy et~al.(2009)Twohy, Coakley~Jr,, and Tahnk}]{twohy2009effect}
Twohy, C.~H., J.~A. Coakley~Jr, and W.~R. Tahnk, 2009: Effect of changes in relative humidity on aerosol scattering near clouds. \textit{Journal of Geophysical Research: Atmospheres}, \textbf{114~(D5)}.

\bibitem[{Twomey(1977)}]{twomey1977influence}
Twomey, S., 1977: The influence of pollution on the shortwave albedo of clouds. \textit{Journal of the atmospheric sciences}, \textbf{34~(7)}, 1149--1152.

\bibitem[{Wetherald and Manabe(1988)Wetherald, and Manabe}]{CloudFeedbackProcessesinaGeneralCirculationModel}
Wetherald, R.~T., and S.~Manabe, 1988: Cloud feedback processes in a general circulation model. \textit{Journal of Atmospheric Sciences}, \textbf{45~(8)}, 1397 -- 1416, \doi{10.1175/1520-0469(1988)045<1397:CFPIAG>2.0.CO;2}, \urlprefix\url{https://journals.ametsoc.org/view/journals/atsc/45/8/1520-0469_1988_045_1397_cfpiag_2_0_co_2.xml}.

\bibitem[{Xue et~al.(2010)Xue, Teller, Rasmussen, Geresdi,, and Pan}]{xue2010effects}
Xue, L., A.~Teller, R.~Rasmussen, I.~Geresdi, and Z.~Pan, 2010: Effects of aerosol solubility and regeneration on warm-phase orographic clouds and precipitation simulated by a detailed bin microphysical scheme. \textit{Journal of the atmospheric sciences}, \textbf{67~(10)}, 3336--3354.

\bibitem[{Yin et~al.(2000)Yin, Levin, Reisin,, and Tzivion}]{yin2000effects}
Yin, Y., Z.~Levin, T.~G. Reisin, and S.~Tzivion, 2000: The effects of giant cloud condensation nuclei on the development of precipitation in convective clouds—a numerical study. \textit{Atmospheric research}, \textbf{53~(1-3)}, 91--116.

\end{thebibliography}

\end{document}